# Semi-Empirical Shadow Molecular Dynamics: A PyTorch implementation

**Authors**

Maksim Kulichenko,[1]* Kipton Barros,[1,2] Nicholas Lubbers,[3] Nikita Fedik,[1,2] Guoqing Zhou,[4] Sergei Tretiak,[1,2,5] Benjamin Nebgen,[1] Anders M. N. Niklasson[1]*

[1]Theoretical Division, Los Alamos National Laboratory, Los Alamos, New Mexico 87545, United States

[2]Center for Nonlinear Studies, Los Alamos National Laboratory, Los Alamos, New Mexico 87545, United States

[3]Computer, Computational, and Statistical Sciences Division, Los Alamos National Laboratory, Los Alamos, New Mexico 87545, United States

[4]Nvidia Corporation

[5]Center for Integrated Nanotechnologies, Los Alamos National Laboratory, Los Alamos, New Mexico 87545, United States

* Corresponding authors: maxim@lanl.gov, amn@lanl.gov

## Abstract

Extended Lagrangian Born-Oppenheimer molecular dynamics (XL-BOMD) in its most recent shadow potential energy version has been implemented in the semiempirical PyTorch-based software PySeQM. The implementation includes finite electronic temperatures, canonical density matrix perturbation theory, and an adaptive Krylov Subspace Approximation for the integration of the electronic equations of motion within the XL-BOMB approach (KSA-XL-BOMD). The PyTorch implementation leverages the use of GPU and machine learning hardware accelerators for the simulations. The new XL-BOMD formulation allows studying more challenging chemical systems with charge instabilities and low electronic energy gaps. Current public release of PySeQM continues our development of modular architecture for large-scale simulations employing semiempirical quantum mechanical treatment. Applied to molecular dynamics simulation of 840 carbon atoms, one integration time step executes in *4 seconds* on a single Nvidia RTX A6000 GPU.

## Introduction

Atomic-scale simulations play a key role in studying chemical processes. High-level quantum-mechanical (QM) methods, such as density functional theory[1] (DFT) and coupled cluster,[2] can provide chemical-level accuracies but at costs which scale in range $O(N^3)$-$O(N^7)$ with the system size, *N*. In practice, molecular dynamics (MD) simulations with full QM forces are typically limited to systems of a couple hundred atoms or less. As an alternative, semiempirical quantum mechanics (SEQM) methods are widely used to model large atomistic systems and run long-time simulations.[3,4] The core of SEQM methods is a reduced-order Hamiltonian in which some elements are replaced by empirical parameters fit to experimental or *ab initio* data. By design, the reduced-order quantum-mechanical model significantly reduces computational burden, but the required introduction of static parameters also compromises accuracy and generality.

The recently introduced PySeQM[5] package supports a variety of semiempirical Hamiltonians such as AM1,[6] MNDO,[7] and PM3[8] models. Written in PyTorch,[9] PySeQM computes interatomic forces via automatic differentiation and supports a batch mode which, in turn, enables efficient simulation of multiple (i.e., in parallel) large molecules with semiempirical Hamiltonians on Graphics Processing Units (GPUs). The latter feature is particularly useful when propagating an ensemble of trajectories, a task frequently required by MD modeling. With automatic differentiation and batching, PySeQM also enables efficient calculation of Hessian matrices which are used for normal modes and infrared (IR) spectra analyses. Various groups have shown that replacing static parameters in reduced-order Hamiltonians with machine-learned (ML) dynamically responsive parameters inferred from the local environment can greatly improve the accuracy of reduced order models.[10,11] Particularly, this approach helped achieve near DFT level of accuracy for semiempirical methods in PySeQM.[12]

Nevertheless, iterative self-consistent field (SCF) optimization of the density matrix (DM) is a significant bottleneck in direct Born-Oppenheimer molecular dynamics (BOMD), even in SEQM methods. Unless the electronic ground state is tightly converged in each time step, the forces will not be conservative, which often leads to large energy drift rendering results physically invalid. To address this issue and further speed up semiempirical MD simulations, PySeQM supports the extended Lagrangian Born Oppenheimer molecular dynamics (XL-BOMD) for accelerated MD simulations which avoids the expensive SCF algorithm.[13,14] This is done by introducing additional extended electronic degrees of freedom, in the spirit of Car-Parrinello MD,[15] and propagating them along with the nuclear motion. This technique preserves time-reversal symmetry and avoids the problem of a systematic total energy drift in MD simulations. This approach resolves a common problem in regular direct BOMD simulations, where it is difficult to achieve sufficient SCF convergence. Although the original XL-BOMD formalism yields significant benefits, there is still room for improvement: XL-BOMD may still require a few SCF iterations per time-step, particularly, in simulations of chemical systems with a small or vanishing HOMO-LUMO gap. Furthermore, the size of the Verlet integration timestep for a traditional XL-BOMD simulation may be somewhat smaller than the time step for the corresponding classical MD simulation.

The main goal of this article is to adapt, implement, and demonstrate the most recent shadow potential formulation of XL-BOMD for simulations using the semiempirical high-performance PySeQM package. In its current implementations, the XL-BOMD uses the single-particle density matrix as a dynamical tensor variable for the propagated electronic degrees of freedom.[16] This formulation is combined with a Krylov Subspace Approximation for the integration of the electronic equations of motion in XL-BOMB (KSA-XL-BOMD)[14,16,17], which now is implemented in the PySeQM code. The integration scheme is based on a tunable, low-rank approximation of a fourth-order kernel, $\mathcal{K}$. The latter determines the metric tensor, $\mathcal{T} \equiv \mathcal{K}^T \mathcal{K}$, used in the extended harmonic oscillator of the Lagrangian that generates the dynamics of the electronic degrees of freedom. The new method gives a highly accurate approximation of the potential energy surface (PES) of direct BOMD, while being significantly faster. No iterative SCF optimization is required prior to the force evaluations, which provides a significant acceleration over regular direct BOMD being a particular advantage for simulations of charge sensitive and low-gap systems. Furthermore, an additional advantage of KSA-XL-BOMD over the original XL-BOMD is its improved stability with respect to the size of the MD time-step with an improved long-term stability. The new scheme also brings an ability to handle more challenging systems with small HOMO-LUMO gaps by a modest additional computational overhead.

This paper is structured as follows. In Section 2 we discuss theoretical details of KSA-XL-BOMD interfaced with semiempirical models in PySeQM. In Section 3, we present results of MD simulations performed using PySeQM MD and electronic structure engines on a GPU. We compare performance and timings of BOMD/XL-BOMD/KSA-XL-BOMD by testing these methods on molecules with a size of up to 840 atoms, including a doubly protonated cyclodecapeptide Gramicidin S and charged fullerenes. Finally, we summarize results in Section 4.

# Methods

The Methods section is divided into 6 subsections. First, we briefly outline the direct BOMD method in subsection 1. Then, we review XL-BOMD formalism in subsection 2 followed by the notes on the variational optimization of the density matrix in subsection 3 and equations of motion in XL-BOMD in subsection 4. The integration of the nuclear and electronic degrees is discussed in subsection 5. The integration of electronic degrees of freedom via Krylov Subspace Approximation (KSA) is introduced in subsection 6.

## 1. Direct Born–Oppenheimer Molecular Dynamics

Consider direct BOMD based on a spin-restricted Hartree–Fock approximation[18,19] that has been generalized to account for fractional occupation numbers and the electronic free energy at finite electronic temperatures.[20] The potential energy surface, $U(\mathbf{R})$, can be defined through a constrained minimization of a nonlinear density-matrix expression of the free energy, where

$$U(\mathbf{R}) = \min_{\mathbf{D}} \{2\text{Tr}[\mathbf{hD}] + \text{Tr}[\mathbf{DG}(\mathbf{D})] - 2T_e\mathcal{S}(\mathbf{f})\} + V_{nn}(\mathbf{R}) \tag{1}$$

under the constraints that

$$\begin{aligned}\mathbf{D} &= \sum_i f_i \mathbf{C}_i \mathbf{C}_i^T \\ \mathbf{C}_i^T \mathbf{S} \mathbf{C}_j &= \delta_{i,j} \\ \sum_i f_i &= N_{\text{occ}}\end{aligned} \tag{2}$$

Here, $\mathbf{R} = \{R_I\}$ are the nuclear coordinates; $\mathbf{D}$ is the single-particle density matrix; $\mathbf{S}$ is the atomistic basis-set overlap matrix; $\mathbf{h}$ is the single-electron Hamiltonian matrix; $N_{\text{occ}}$ is the number of occupied states (for simplicity we assume double occupancy representing a common closed shell case); $f_i \in [0, 1]$ are the fractional occupation numbers; $\{\mathbf{C}_i\}$ is a set of eigenvectors corresponding to the coefficients of occupied molecular orbitals; $V_{nn}$ is the ion–ion repulsive term; and $\mathbf{G}(\mathbf{D})$ is the electron-electron interaction matrix,

$$\mathbf{G}(\mathbf{D}) = 2\mathbf{J}(\mathbf{D}) - \mathbf{K}(\mathbf{D}) \tag{3}$$

where $\mathbf{J}(\mathbf{D})$ and $\mathbf{K}(\mathbf{D})$ are Coulomb and exchange matrices, respectively.
$T_e$ is the electronic temperature, and $\mathcal{S}(\mathbf{f})$ is the electronic entropy contribution,

$$\mathcal{S}(\mathbf{f}) = -k_B \sum_i \left( f_i \ln(f_i) + (1-f_i)\ln(1-f_i) \right) \tag{4}$$

where $k_B$ is Boltzmann's constant. The ground state density matrix $\mathbf{D}_{\min}$, which defines the Born-Oppenheimer PES, $U(\mathbf{R})$ in eq. 1, can be calculated from the self-consistent solution of the generalized nonlinear algebraic eigenvalue equation in the mean field approximation,

$$\mathbf{F}(\mathbf{D})\mathbf{C}_i = \mathbf{S}\mathbf{C}_i \epsilon_i \tag{5}$$

where $\mathbf{F}(\mathbf{D})$ is Fock matrix defined as

$$\mathbf{F}(\mathbf{D}) = \mathbf{h} + \mathbf{G}(\mathbf{D}) \tag{6}$$

MNDO, AM1, and PM3 semiempirical methods are based on the Neglect of Diatomic Differential Overlap (NDDO) approximation, i.e. where $\mathbf{S}=\mathbf{I}$. However, for the sake of generality, we keep $\mathbf{S}$ in our formulation. The density matrix, $\mathbf{D}$, is then defined as a weighted outer product of the eigenvectors (eq. 2) with the Fermi occupation factors (the fractional occupation numbers)

$$f_i = \left( e^{\beta(\epsilon_i - \mu_0)} + 1 \right)^{-1} \tag{7}$$

Here $\beta = 1/(k_B T_e)$ is the inverse temperature. The chemical potential, $\mu_0$, is chosen such that

$$\sum_i f_i = N_{\text{occ}} \tag{8}$$

The solution of $\mathbf{D}_{\min}$ in the constrained, nonlinear, variational formulation in eq. 1, given through eqs. 5–8, can be found via an iterative self-consistent-field optimization procedure. This iterative optimization is usually computationally expensive and, in practice, never exact.

Once a sufficiently well-converged ground state density matrix, $\mathbf{D}_{\min}$, is found, the interatomic forces can be calculated using the Hellmann–Feynman theorem. The molecular trajectories can then be generated by integrating Newton's equation of motion,

$$M_I \ddot{R}_I = -\frac{\partial U(\mathbf{R})}{\partial R_I} \tag{9}$$

where $M_I$ are the atomic masses. Typically, analytic gradient techniques compute gradients (forces) efficiently. For example, interatomic forces in PySeQM are computed via reverse-mode automatic differentiation,[9] which efficiently computes the gradient of a quantity with respect to all parameters.

Finally, the constant of motion is defined by the Born–Oppenheimer Hamiltonian,

$$\mathcal{H}_{\text{BO}} = \frac{1}{2} \sum_I M_I \dot{R}_I^2 + U(\mathbf{R}) \tag{10}$$

## 2. Extended Lagrangian Born–Oppenheimer Molecular Dynamics

By extrapolating the ground state density matrix from previous time steps, it is possible to reduce the computational overhead of the iterative ground state optimization required in eq. 1. However, because the variational ground state optimization is approximate and never complete (i.e., $\partial U(\mathbf{R})/\partial \mathbf{D} \approx 0$), the forces evaluated using the Hellmann–Feynman theorem are never exactly conservative. This causes a systematic total energy drift because the fictitious propagation of electronic degrees of freedom breaks time-reversal symmetry.[21–23] Alternatively, one can restart the ground state optimization in each time step from overlapping atomic densities. This approach preserves the time-reversal symmetry and avoids a systematic energy drift, but there is a large computational expense to converge the SCF to very high precision. Extended Lagrangian Born–Oppenheimer molecular dynamics (XL-BOMD)[13,14,16,24,25] is a framework developed to avoid these shortcomings.

The key idea behind XL-BOMD in its most recent shadow potential formulation is based on a backward error analysis or a shadow Hamiltonian approach.[14,26] Rather than calculating approximate forces through a time-consuming iterative ground-state optimization for an "exact" Born–Oppenheimer potential energy surface (PES), we can do the opposite. Namely, exact forces can be calculated in a fast SCF-free way, but for an approximate *shadow* PES. In this way we can reduce the computational cost and at the same time restore a consistency between the calculated forces and the underlying shadow potential.

Density-matrix based XL-BOMD is given in terms of an extended Lagrangian formulation of the dynamics using four approximations: (1) The nonlinear density matrix energy expression that is minimized in eq. 1 is linearized around some approximate density matrix, $\mathbf{P}$, which is assumed to be close to the exact ground state, $\mathbf{D}_{\text{min}}$. The constrained variational optimization of this linearized energy expression provides the $\mathbf{P}$-dependent ground state density matrix, $\mathbf{D}_0[\mathbf{P}]$, and shadow potential energy surface, $\mathcal{U}(\mathbf{R}, \mathbf{P})$, which can be determined by a single SCF-free step without the approximate iterative ground state optimization. (2) The approximate density matrix, $\mathbf{P}$, around which the linearization is performed, is included as an extended dynamical tensor variable for a fictitious electronic degree of freedom that evolves through a harmonic oscillator that is centered around the best available approximation to the exact ground state density matrix, $\mathbf{D}_{\text{min}}$. In this way $\mathbf{P}$ will remain close to the optimized ground state density. (3) The harmonic potential centered around the best available approximation of the ground state density matrix is given by a generalized square-norm of the residual matrix function, $\mathbf{D}_0[\mathbf{P}] - \mathbf{P}$, using a metric tensor, $\mathcal{T} \equiv \mathcal{K}^T \mathcal{K}$, where $\mathcal{K}$ is the inverse Jacobian of the residual matrix function, $\mathbf{D}_0[\mathbf{P}] -$

**P**. (4) The Euler–Lagrange equations of motion are then derived in an adiabatic limit, where the frequency of the harmonic oscillator extension is assumed to be high compared to the highest frequency of the nuclear degrees of freedom and the fictitious electron mass parameter goes to zero. This classical adiabatic limit corresponds to a Born–Oppenheimer-like approximation for the extended classical electronic degrees of freedom and leads to a pair of coupled equations of motion for the nuclear and the electronic degrees of freedom. A similar "mass-zero" limit can also be imposed onto the equations of motion in Car-Parrinello molecular dynamics using Lagrange multipliers.[27,28]

The extended Lagrangian for spin-restricted thermal Hartree–Fock theory, which includes fractional occupation numbers at finite electronic temperatures, is defined by

$$\mathcal{L}(\mathbf{R}, \dot{\mathbf{R}}, \mathbf{X}, \dot{\mathbf{X}}) = \frac{1}{2} \sum_I M_I \dot{R}_I^2 - \mathcal{U}(\mathbf{R}, \mathbf{X}) + \frac{\mu}{2} \text{Tr}\left[|\dot{\mathbf{X}}|^2\right] - \frac{\mu \omega^2}{2} \text{Tr}\left[(\mathbf{D_0}[\mathbf{X}]\mathbf{S} - \mathbf{X})^T \mathcal{T} (\mathbf{D_0}[\mathbf{X}]\mathbf{S} - \mathbf{X})\right] \quad (11)$$

The dynamical matrix variables $\mathbf{X}$ represent the extended electronic degrees of freedom. $\mathcal{U}(\mathbf{R}, \mathbf{X})$ is the shadow potential for the electronic free energy at some electronic temperature $T_e$ that approximates the corresponding exact Born–Oppenheimer free energy surface; $\mathcal{T} \equiv \mathcal{K}^T \mathcal{K}$ is the metric tensor; $\mu$ is a fictitious electronic mass parameter; and $\omega$ is the frequency of the harmonic oscillator extension. $\mathbf{D_0}[\mathbf{X}]$ is the $\mathbf{X}$-dependent density matrix given by the constrained variational minimization of a linearized density matrix function for the electronic energy,

$$\mathbf{D_0}[\mathbf{X}] = \arg\min_{\mathbf{D}} \left\{ 2\text{Tr}[\mathbf{hD}] + \text{Tr}\left[(2\mathbf{D} - \mathbf{X}\mathbf{S}^{-1}) \mathbf{G} (\mathbf{X}\mathbf{S}^{-1})\right] - 2T_e \mathcal{S}(\mathbf{f}) \right\} \quad (12)$$

where it is assumed that $\mathbf{P} \equiv \mathbf{X}\mathbf{S}^{-1} \approx \mathbf{D_0}[\mathbf{X}]$ is an approximate ground state density matrix. Because of the linearization, this minimization can be performed exactly in a single step, without any iterative SCF optimization. This both improves the accuracy and reduces the cost. The shadow potential energy surface, $\mathcal{U}(\mathbf{R}, \mathbf{X})$, in eq. 11 is then given by

$$\mathcal{U}(\mathbf{R}, \mathbf{X}) = 2\,\text{Tr}[\mathbf{hD_0}[\mathbf{X}]] + \text{Tr}\left[(2\mathbf{D_0}[\mathbf{X}] - \mathbf{X}\mathbf{S}^{-1}) \mathbf{G} (\mathbf{X}\mathbf{S}^{-1})\right] - T_e \mathcal{S}(\mathbf{f}) + V_{nn}(\mathbf{R}) \quad (13)$$

which is a close approximation to the fully relaxed ground state Born–Oppenheimer potential energy surface if the magnitude of the residual matrix function, $\mathbf{D_0}[\mathbf{X}]\mathbf{S} - \mathbf{X}$, is small, for example, as estimated by a Frobenius norm.

The metric tensor $\mathcal{K}$ in the harmonic oscillator extension is chosen such that the dynamical density matrix variable, $\mathbf{P}$, for the extended electronic degrees of freedom, oscillates around a much closer approximation to the exact Born–Oppenheimer ground state compared to the variationally optimized solution of the linearized energy functional, $\mathbf{D_0}[\mathbf{P}]$. This improves the stability and the adiabatic decoupling between the nuclear and electronic degrees of freedom.

The proposed formulation requires a fourth-order metric tensor, $\mathcal{T}$, that performs a mapping between matrices. As in previous density matrix formulations of XL-BOMD,[16,29] we will use a general atomic-orbital representation of the extended electronic degrees of freedom with a dynamical density matrix variable $\mathbf{X} = \mathbf{PS}$ instead of the regular density matrix $\mathbf{P} = \mathbf{X}\mathbf{S}^{-1}$ or its orthogonalized form, $\mathbf{P}^\perp$.

Since the overlap matrix is approximated as an identity (**S**=**I**) matrix in semiempirical methods implemented in PySeQM, the relation **X** = **P** is valid**.** However, we will keep the **X** notation for the purposes of generality.

The expression for the harmonic oscillator of the extended Lagrangian in eq. 11 includes a metric tensor,

$$\mathcal{T} \equiv \mathcal{K}^T \mathcal{K} \tag{14}$$

where $\mathcal{K}$ is a kernel that acts as a fourth-order tensor, which performs mappings between matrices. This kernel, $\mathcal{K}$, is defined from the inverse of the Jacobian, $\mathcal{J}$, of the residual matrix function, where

$$\mathcal{J}_{ij,kl} = \frac{\partial(\{\mathbf{D_0}[\mathbf{X}]\mathbf{S}\}_{ij} - X_{ij})}{\partial X_{kl}} \tag{15}$$

and

$$\mathcal{K} = \mathcal{J}^{-1} \tag{16}$$

We will only deal with a low-rank approximations of $\mathcal{K}$ acting on the residual matrix function. Low-rank approximations are necessary to avoid a large computational overhead.

## 3. Variational Optimization of the Density Matrix

The main computational cost of XL-BOMD is the constrained variational optimization of the density matrix **D** in eq. 12. However, the cost is drastically reduced compared to the nonlinear minimization required in direct Born–Oppenheimer molecular dynamics in eq. 1.

If the approximate Fockian in a nonorthogonal atomic-orbital representation is given by

$$\mathbf{F} \equiv \mathbf{F}(\mathbf{X}) = \mathbf{h} + \mathbf{G}(\mathbf{X}\mathbf{S}^{-1}) \tag{17}$$

with the orthogonalized matrix representation,

$$\mathbf{F}_0^\perp = \mathbf{Z}^T \mathbf{F} \mathbf{Z} \tag{18}$$

then the constrained density matrix minimization in eq. 12 can be performed by first calculating

$$\mathbf{D_0}^\perp[\mathbf{X}] = \left(e^{\beta(\mathbf{F}_0^\perp - \mu_0 \mathbf{I})} + \mathbf{I}\right)^{-1} \tag{19}$$

where $\mu_0$ is the chemical potential set such that Tr[$\mathbf{D_0}^\perp[\mathbf{X}]$] = $N_{occ}$, followed by the transform back to the nonorthogonal atomic orbital representation, where

$$\mathbf{D_0}[\mathbf{X}] = \mathbf{Z}\mathbf{D_0}^\perp[\mathbf{X}]\mathbf{Z}^T \tag{20}$$

Here **Z** is a matrix, which in the symmetric case corresponds to a Löwdin orthogonalization, where **Z**=**S**$^{-\frac{1}{2}}$. In the general case, we can use any matrix **Z** that fulfills the condition

$$\mathbf{Z}^T\mathbf{S}\mathbf{Z} = \mathbf{I} \tag{21}$$

The constrained optimization for the ground state solution $\mathbf{D_0[X]}$ in eq. 12 is thus given without any iterative self-consistent field optimization, because the matrix functional in eq. 12 is linear in $\mathbf{D}$. The expensive nonlinear self-consistent field problem in regular Born–Oppenheimer molecular dynamics, which requires an iterative solution, has thus been removed. $\mathbf{D_0[X]}$ is the exact and variationally stationary solution of the shadow potential energy surface, $\mathcal{U}(\mathbf{R}, \mathbf{X})$, in eq. 13, in the same way as the exact $\mathbf{D}_{\text{min}}$ is the variationally stationary solution for the regular Born–Oppenheimer potential energy surface, $U(\mathbf{R})$, in eq. 1. This simplifies the calculation of interatomic forces that are consistent with the shadow potential energy surface, and we can avoid contributions from terms including $\partial \mathbf{D_0[X]}/\partial R_I$, because

$$\left.\frac{\partial \mathcal{U}(\mathbf{R}, \mathbf{X})}{\partial \mathbf{D_0[X]}} \frac{\partial \mathbf{D_0[X]}}{\partial R_I}\right|_{\mathbf{D_0}} = 0 \tag{22}$$

As long as $\mathbf{XS}^{-1}$ is a reasonably close approximation to the exact self-consistent ground state density, $\mathbf{D}_{\text{min}}$, i.e., as long as the residual matrix function $\mathbf{D_0[X]S} - \mathbf{X}$ is small, the shadow potential energy surface, $\mathcal{U}(\mathbf{R}, \mathbf{X})$, in eq. 13, is close to the exact Born–Oppenheimer potential, $U(\mathbf{R})$.[30] This can be understood from the fact that the difference between the shadow potential and the regular BO potentials scales with the second order of the DM residual,[14] i. e.

$$\mathcal{U}(\mathbf{R}, \mathbf{X}) = U(\mathbf{R}) + \mathcal{O}(||\mathbf{D_0[X]S} - \mathbf{X}||_F^2) \tag{23}$$

## 4. Equations of Motion

In the derivation of the equations of motion from the Euler–Lagrange equations for the extended Lagrangian in eq. 11, we apply an adiabatic approximation that separates the motion between the nuclear and the extended electronic degrees of freedom. This is consistent with and a direct classical analogue to the original Born-Oppenheimer approximation, which is used to separate the motion between the slow nuclear and fast electronic degrees of freedom. In our classical adiabatic approximation, we assume that $\omega$, which determines the frequency of the extended electronic degrees of freedom, $\mathbf{X}(t)$ and $\dot{\mathbf{X}}(t)$, is large compared to some highest frequency, $\Omega$, of the nuclear degrees of freedom, $\mathbf{R}(t)$ and $\dot{\mathbf{R}}(t)$. See ref [16] for more details. The equations of motion for XL-BOMD in this adiabatic mass-zero limit are then given by

$$M_I \ddot{R}_I = \left.\frac{\partial \mathcal{U}(\mathbf{R}, \mathbf{X})}{\partial R_I}\right|_{\mathbf{X}} \tag{24}$$

for the nuclear degrees of freedom and

$$\ddot{\mathbf{X}} = -\omega^2 \mathcal{K} \left(\mathbf{D_0[X]S} - \mathbf{X}\right) \tag{25}$$

for the electronic degrees of freedom. The partial derivatives in eq. 24 are calculated with respect to a constant $\mathbf{X}$, because $\mathbf{X}$ is a dynamical variable. In this way the simplicity similar to the force term in regular direct BOMD for the exact fully optimized ground state solution $\mathbf{D}_{min}$ is recovered, even if $\mathbf{X}$ is not the exact ground state solution. The corresponding constant of motion is given by,

$$\mathcal{H}_{\text{XL-BOMD}} = \frac{1}{2}\sum_I M_I \dot{R}_I^2 + \mathcal{U}(\mathbf{R}, \mathbf{X})$$

$$= \frac{1}{2}\sum_I M_I \dot{R}_I^2 + 2\text{Tr}\left[\mathbf{h}\mathbf{D_0}[\mathbf{X}]\right] + \text{Tr}\left[\left(2\mathbf{D_0}[\mathbf{X}] - \mathbf{X}\mathbf{S}^{-1}\right)\mathbf{G}\left(\mathbf{X}\mathbf{S}^{-1}\right)\right] - T_e \mathcal{S}(\mathbf{f}) + V_{nn}(\mathbf{R})$$
(26)

Equations 24 and 25 together with the constant of motion in eq. 26 are the three central equations that govern the dynamics of XL-BOMD for thermal Hartree–Fock theory.

## 5. Integrating XL-BOMD Using a Modified Verlet Scheme with Damping

For the integration of the combined nuclear and electronic degrees of freedom in eqs. 24 and 25 we use a modified leapfrog velocity Verlet scheme[24,31,32] that includes an additional dissipative term in the integration of the extended electronic degrees of freedom. This additional term breaks the time-reversal symmetry to some chosen higher odd-order in the integration time step, $\delta t$, which dampens the accumulation of numerical noise that otherwise could cause instabilities in a perfectly reversible integration. In this way the evolution of the electronic degrees of freedom stays synchronized to the dynamics of the nuclear motion. The modified leapfrog velocity Verlet integration scheme for the integration of the nuclear and electronic degrees of freedom is given by

$$\dot{\mathbf{R}}(t + \frac{\delta t}{2}) = \dot{\mathbf{R}}(t) + \frac{\delta t}{2}\ddot{\mathbf{R}}(t) \tag{27}$$

$$\mathbf{R}(t + \delta t) = \mathbf{R}(t) + \delta t \dot{\mathbf{R}}(t + \frac{\delta t}{2}) \tag{28}$$

$$\mathbf{X}(t + \delta t) = 2\mathbf{X}(t) - \mathbf{X}(t - \delta t) + \kappa \ddot{\mathbf{X}}(t) + \alpha \sum_{k=0}^{k_{max}} c_k \mathbf{X}(t - k\delta t) \tag{29}$$

$$\dot{\mathbf{R}}(t + \delta t) = \dot{\mathbf{R}}(t + \frac{\delta t}{2}) + \frac{\delta t}{2}\ddot{\mathbf{R}}(t + \delta t) \tag{30}$$

The last term in the integration of $\mathbf{X}(t)$ is the additional damping term, where the coefficients, $\alpha$ and $\{c_k\}_{k=0}^{k_{max}}$, as well as a dimensionless constant $\kappa = \delta t^2 \omega^2$, have been optimized for various values of $k_{max}$ and are given in ref [24]. Here, we use kmax=6 but it was shown that any value between 3 and 9 gives good results.[32] In the initial time step $\mathbf{X}(t_0 - k\delta t)$ for $k = 0, 1, ..., k_{max}$ are all set to the fully converged regular Born–Oppenheimer ground state density, $\mathbf{D}_{min}$, times the overlap matrix $\mathbf{S}$ (which is approximated as an identity matrix in PySeQM); i.e., at $t_0$ we set $\mathbf{X}(t_0 - k\delta t) = \mathbf{D}_{min}\mathbf{S}$ for $k = 0, 1, ..., k_{max}$. $\ddot{\mathbf{X}}$ is given by the eq 25 and discussed in more detail in the next subsection. A reasonably well-converged iterative self-consistent field optimization is thus required, but only in the first initial time step. The modified Verlet integration scheme works similar to a Langevin dynamics, but where the stochastic term is generated by the intrinsic numerical noise of the system instead of an external random number generator and where the dissipation is given by a higher-order time-derivative term instead of a first-order velocity-driven friction term. In general, this solution works well without any significant drift in the constant of motion on time scales relevant for quantum-based Born–Oppenheimer molecular dynamics. A detailed discussion on time scales and energy drifts in XL-BOMD can be found in reference [14]. Some alternative integration methods have also been proposed by Teresa Head-Gordon and co-workers in applications to flexible charge models.[33–35]

## 6. Krylov Subspace Approximation of the Inverse Jacobian Kernel

The integration of the electronic degrees of freedom in eq. 29 requires the calculation of the second-order time derivative of $\mathbf{X}(t)$, which is given by the electronic equations of motion in eq. 25. This equation includes the kernel, $\mathcal{K}$ that is given from the inverse Jacobian of the residual matrix function ($\mathbf{D}[\mathbf{X}]\mathbf{S}$-$\mathbf{X}$) using eqs. 15 and 16. A technique to perform a tunable and adaptive approximation of the kernel for the integration of the electronic degrees of freedom in XL-BOMD was developed,[17] but for a residual function $\mathbf{f}(\mathbf{n})=(\mathbf{q}[\mathbf{n}]-\mathbf{n})$. That method was derived from an expression of the Jacobian of $\mathbf{f}(\mathbf{n})$, which is based on a set of general directional derivatives,

$$\mathbf{f}_{\mathbf{v}_i} = \left.\frac{\partial \mathbf{f}(\mathbf{n} + \lambda \mathbf{v}_i)}{\partial \lambda}\right|_{\lambda=0} \tag{31}$$

along directions $\mathbf{v}_i$ instead of partial derivatives, $\partial \mathbf{f}(\mathbf{n})/\partial n_i$, with respect to the individual components of $\mathbf{n}$. The Jacobian, $\mathbf{J}$, in this case, can then be approximated by a rank-$m$ expression,

$$\mathbf{J} \approx \sum_{i,j=1}^{m} \mathbf{f}_{\mathbf{v}_i} L_{ij} \mathbf{v}_j^T \tag{32}$$

where $L = O^{-1}$, which is the inverse overlap with matrix elements, $O_{ij} = \mathbf{v}_i^T \mathbf{v}_j$. The directional derivatives are calculated with quantum perturbation theory, and the directions $\{\mathbf{v}_i\}_{i=1}^{m}$ are chosen from a Krylov subspace approximation.[17] The kernel is then determined by a pseudoinverse of the low-rank Jacobian. In combination with preconditioning, this provides a rapidly converging low-rank approximation of the kernel in the integration of the electronic degrees of freedom.

Here we adapt the previous approximation method of the Jacobian and its pseudoinverse to the residual matrix function, $(\mathbf{D_0}[\mathbf{X}]\mathbf{S} - \mathbf{X})$. We start by rewriting eq. 25 in a more general but equivalent form, using a preconditioner, $\mathcal{K}_0$, i.e., where

$$\ddot{\mathbf{X}} = -\omega^2 (\mathcal{K}_0 \mathcal{J})^{-1} \mathcal{K}_0 \left(\mathbf{D_0}[\mathbf{X}]\mathbf{S} - \mathbf{X}\right) \tag{33}$$

In the ideal case we assume that $\mathcal{K}_0 \mathcal{J} \approx \mathcal{I}$, where $\mathcal{I}$ is the fourth-order identity tensor. We will then try to find a low-rank approximation of how the preconditioned kernel, $(\mathcal{K}_0 \mathcal{J})^{-1}$, acts on the modified residual, $\mathcal{K}_0 \left(\mathbf{D_0}[\mathbf{X}]\mathbf{S} - \mathbf{X}\right)$. For the second-order kernels, it is possible to construct efficient preconditioners by a direct full calculation of the kernel at the initial time step. The latter then can be kept as a constant preconditioner during the molecular dynamics simulation. This preconditioning technique generates rapidly converging low-rank approximations of the kernel. With the fourth-order kernel, this direct method is no longer possible, except for very small molecular systems. We will therefore not pursue the use of preconditioners here, even if they can be constructed in different forms. We will nevertheless keep a $\mathcal{K}_0$ in our discussion for the sake of generality. It will be demonstrated by several examples that we can achieve accurate low-rank approximations even without preconditioner, i.e., when the identity is used as a preconditioner and $\mathcal{K}_0 = \mathcal{I}$. Another difference to the original Krylov subspace approximation is that the metric of the inner product has to be modified. Instead of a vector dot-product, $\langle \mathbf{v}_i, \mathbf{v}_j \rangle = \mathbf{v}_i^T \mathbf{v}_j$, we will use the matrix generalization, $\langle \mathbf{V}_i, \mathbf{V}_j \rangle = \mathrm{Tr}[\mathbf{V}_i^T \mathbf{V}_j]$. The matrix norm, $\|\mathbf{V}_i\| = \sqrt{\langle \mathbf{V}_i, \mathbf{V}_i \rangle} = \sqrt{\mathrm{Tr}[\mathbf{V}_i^T \mathbf{V}_i]}$, then corresponds to the Frobenius norm. The general principle for the original formulation can then be kept. With this matrix generalization of the method in ref [17], the rank-*m* Krylov subspace approximation for how the preconditioned kernel acts on the modified matrix residual function is given by

$$(\mathcal{K}_0 \mathcal{J})^{-1} \mathcal{K}_0 \left( \mathbf{D_0}[\mathbf{X}]\mathbf{S} - \mathbf{X} \right) \approx \sum_{i,j=1}^{m} \mathbf{V}_i M_{ij} \langle \mathbf{W}_j, \mathbf{W}_0 \rangle \quad (34)$$

The algorithm that generates the matrices $\mathbf{V_i}$, $\mathbf{M}$, and $\mathbf{W_i}$, is given in ref [16], where $\mathbf{W_0} = \mathbf{K_0}(\mathbf{D}[\mathbf{X}]\mathbf{S} - \mathbf{X})$. It is an adaptive scheme, where the order *m* of the rank-*m* approximation is tunable by a chosen tolerance level. With this tunable framework of a rank-*m* Krylov Subspace Approximation in XL-BOMD (KSA-XL-BOMD), we can then integrate the electronic degrees of freedom in eq 29 as

$$\mathbf{X}(t + \delta t) = 2\mathbf{X}(t) - \mathbf{X}(t - \delta t) - \delta t^2 \omega^2 \sum_{i,j=1}^{m} \mathbf{V}_i M_{i,j} \langle \mathbf{W}_j, \mathbf{W}_0 \rangle + \alpha \sum_{k=0}^{k_{\max}} c_k \mathbf{X}(t - k\delta t) \quad (35)$$

The calculation of $\mathbf{W}_m$ requires the directional derivatives of the density matrix

$$\partial_\lambda \mathbf{D}[\mathbf{F}_0(\mathbf{X}) + \lambda \mathbf{F}_1(\mathbf{V}_m)]|_{\lambda=0} = \mathbf{Z} \left( \frac{\partial}{\partial \lambda} \mathbf{D}^\perp [\mathbf{F}_0^\perp + \lambda \mathbf{F}_1^\perp (\mathbf{V}_m)] \bigg|_{\lambda=0} \right) \mathbf{Z}^T \quad (36)$$

This is done using a canonical density matrix perturbation theory within thermal Hartree-Fock formalism. See ref [16] for details and the pseudocode.

Finally, the force evaluations are based on

$$\frac{\partial \mathcal{U}(\mathbf{R}, \mathbf{X})}{\partial R_I} = 2\mathrm{Tr}[\mathbf{h}_{R_I} \mathbf{D}] + \mathrm{Tr}[(2\mathbf{D} - \mathbf{P})\mathbf{G}_{R_I}] + \partial V_{nn}/\partial R_I - 2\mathrm{Tr}[\mathbf{Z}\mathbf{Z}^T \mathbf{F}\mathbf{D}\mathbf{S}_{R_I}] \quad (37)$$

and include terms like

$$\mathbf{h}_{R_I} = \frac{\partial \mathbf{h}}{\partial R_I} \quad (38)$$

$$\mathbf{S}_{R_I} = \frac{\partial \mathbf{S}}{\partial R_I} \quad (39)$$

$$\mathbf{G}_{R_I} = \frac{\partial \mathbf{G}(\mathbf{R}, \mathbf{P})}{\partial R_I} \bigg|_{\mathbf{P}} \quad (40)$$

$$f_{\text{Pulay}} = -2\text{Tr}[\mathbf{ZZ}^T\mathbf{FDS}_{R_I}] \tag{41}$$

In PySeQM, the different force terms in eq 37 are computed automatically using reverse-mode automatic differentiation. Since we approximate the overlap matrix $\mathbf{S}$ as an identity matrix, the Pulay force-term[36] $f_{\text{Pulay}}$ vanishes and, again, $\mathbf{P}=\mathbf{X}$.

# Results and Discussion

For the test systems described below, we use the AM1 Hamiltonian and a fixed electronic temperature $T_e$ of 1500K. This value is small enough to have a negligible effect of fractional occupations on the simulations but also large enough to activate the thermal Hartree-Fock formalism for the integration of the electronic equations of motion via Krylov Subspace Approximation. In general, test cases discussed below are not very sensitive to this parameter: the values in range 100 - 4000K exhibit similar behavior.

We apply BOMD at the AM1 level of theory with the convergence criteria of $10^{-6}$ eV as a baseline benchmark. Although the AM1 Hamiltonian is not the most advanced among modern semi-empirical methods,[10,12,37-40] it is a well-established semi-empirical approximation. Its broad adoption makes AM1 a good choice for the current study where our main goal is to introduce, demonstrate and analyze the general KSA-XL-BOMD methodology for semi-empirical quantum chemistry theory. We will also highlight the relative computational performance of KSA-XL-BOMD over the conventional BOMD. This comparison is less sensitive to a particular choice of semi-empirical Hamiltonian.

## Methanol IR Spectrum

With a simple electronic structure, methanol is a good starting point for testing the semi-empirical shadow molecular dynamics presented in this work. A good way to track the general quality of a DM propagation is the calculation of the InfraRed (IR) absorption spectrum via a Fourier transform of a dipole moment autocorrelation function from the MD trajectories, because the dipole moment is directly related to the DM. The overall advantage of MD-derived IR spectrum over simple normal modes analysis is that it provides experiment-like broadening, relative intensities, and is not constrained by a harmonic approximation. It also gives an averaged picture of the dynamics over a broad frequency spectrum.

The structure was initially optimized using a steepest descent optimizer as implemented in PySeQM. Initializing from the optimized structure and randomly assigning velocities, we further run MD with a Langevin thermostat for 350,000 steps with a 0.4 fs time step and 300K ionic temperature. Here, we compare the performance of direct BOMD, XL-BOMD, and rank-3 KSA-XL-BOMD. In the direct BOMD approach, the density matrix is initialized from scratch from overlapping atomic densities at each time step to avoid systematic energy drift.

Figure 1 depicts the IR spectrum of a methanol molecule calculated as a Fourier transform of a dipole moment autocorrelation function. We also provide spectrum derived from the normal mode analysis, which is performed using a semi-numerical approach. That is, forces of 18 (degrees of freedom in methanol molecule) structures with displaced atoms (both back and forward displacements) are calculated using automatic differentiation in a batch mode. Then, the

Hessian matrix is obtained using a finite differences approach. Relative intensities are calculated numerically as a square of a dipole moment change with respect to displacements along vibrational coordinates defined as

$$I = \left(\frac{\partial \mu}{\partial q}\right)^2 \tag{42}$$

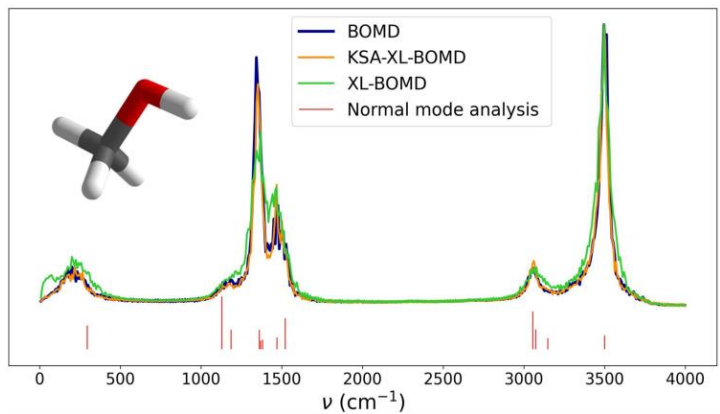

Figure 1. Normalized IR spectra of a methanol molecule (gas phase approximation) calculated as a Fourier transform of a dipole moment autocorrelation function from BOMD (blue), rank-3 KSA-XL-BOMD (orange), and XL-BOMD (green) MD simulations. Red lines represent scaled IR peaks from normal modes analysis (plotted as offset).

The shape of the spectrum derived from the KSA-XL-BOMD simulation is in a near-perfect agreement with the one from the reference tightly converged direct BOMD simulation over the full spectral range as shown in Fig. 1. Meanwhile, the spectrum from XL-BOMD exhibits small but still noticeable deviations. In the 0-500 cm$^{-1}$ region, the spectrum from XL-BOMD displays a significantly larger amount of noise compared to BOMD and KSA-XL-BOMD. In the 1000-1500 cm$^{-1}$ region, both BOMD and KSA-XL-BOMD provide three distinct peaks. Meanwhile, in the XL-BOMD simulation, the small peak at ~1100 almost merges with the major band because of the excessive broadening. In the 3000-3600 cm$^{-1}$ region, there is a near-perfect agreement between BOMD and KSA-XL-BOMD. XL-BOMD also provides a good agreement with direct BOMD, although having a slightly wider broadening.

The RMSE between whole normalized BOMD and XL-BOMD spectra is 0.048 vs. 0.026 for KSA-XL-BOMD. Thus, the example of a methanol molecule demonstrates that introduction of finite electronic temperatures, canonical density matrix perturbation theory, and an adaptive Krylov subspace approximation for the integration of the electronic equations of motion improves the DM dynamics already for a fairly simple molecular system.

Since methanol is a small six-atomic molecule, it is hard to properly estimate the computational overhead of SCF procedures and DM propagation on a GPU with such small density matrices. In next sections, we analyze larger molecules and estimate the relative timing of BOMD/XL-BOMD/KSA-XL-BOMD.

## Gramicidin S Molecular Dynamics

The next test case is focused on the relatively large and biologically important molecule Gramicidin S (Figure 2.a).[41] Gramicidin S is a cyclodecapeptide, an antibiotic that is effective against some gram-positive and gram-negative bacteria as well as some fungi. Here, we perform MD simulations of doubly protonated Gramicidin S, $[GS + 2H]^{2+}$, a 176-atom system[42–44] with molecular formula $C_{60}N_{12}O_{10}H_{94}$, in the NVE ensemble. Initialized at 300K, the statistical temperature kept close to 150K during the simulation.

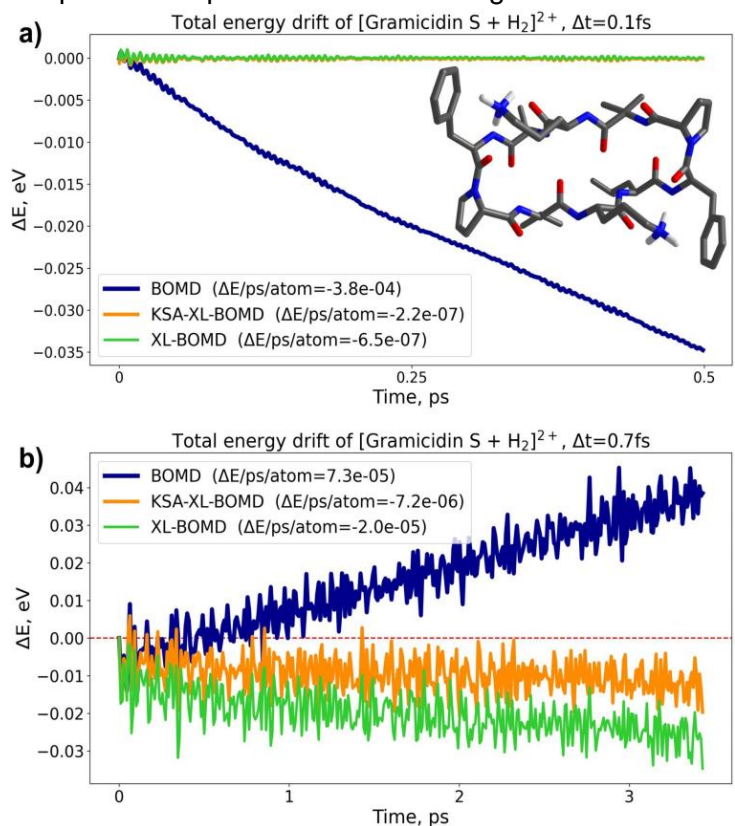

Figure 2. Total energy in MD trajectories of $[GS + 2H]^{2+}$ system with a) dt=0.1 fs and b) dt=0.7 fs. NVE ensemble. In $[GS + 2H]^{2+}$ inset, carbon atoms are colored in gray, oxygen in red, nitrogen in blue, hydrogen in white. Hydrogen atoms are hidden except the ones bonded to protonated nitrogens. The statistical temperature fluctuates around 150K.

It is of key importance that quantum-based MD simulations conserve total energy for long-time simulations in the NVE ensemble. Due to the finite SCF convergence combined with an initial guess extrapolated from the previous MD time steps, most BOMD simulations exhibit systematic total energy drift over the MD trajectory. For example, Figure 2.a (blue line) shows a BOMD simulation with the time step of 0.1 fs where at each time step the DM from the previous step is used as an initial guess and the SCF ground state is considered to be converged when no changes in electronic energy larger than $10^{-6}$ eV observed. This results in a systematic $-3.8\times10^{-4}$ eV/ps/atom drift of the total energy. XL-BOMD (Figure 2.a, green line), as well as KSA-XL-BOMD (orange line), successfully address this problem having statistically negligible drifts.

Figure 2.b illustrates an important improvement of KSA-XL-BOMD over the original XL-BOMD: the new method is more robust with respect to larger time steps. With the time step of 0.7 fs, BOMD and XL-BOMD have drift values of $7.3\times10^{-5}$ and $-2.0\times10^{-5}$ eV/ps/atom, respectively.

Although KSA-XL-BOMD also starts drifting at this time step size, its energy drift is 2.8 times slower than that in XL-BOMD being $-7.2\times10^{-6}$ eV/ps/atom.

Figure 3.a depicts the total energy in a 40 ps MD simulation using Langevin thermostat with a 0.2 fs time step and 300K ionic temperature. Note that for this NVT simulation the constant of motion will not be conserved. The velocities are assigned with the same randomization seed for each simulation. However, as can be seen in Figure 3.a, there is a noticeable discrepancy in the total energies between the BOMD and XL-BOMD methods by the end of the simulation which also results in HOMO-LUMO underestimation (Figure 3.b). Meanwhile, in the direct BOMD and KSA-XL-BOMD simulations the total energies stay correlated till the end of the trajectory. This difference suggests that the electronic degrees of freedom in XL-BOMD exhibit more noise and therefore more rapidly decorrelate from the BOMD baseline by the end of the simulation. Further, the direct BOMD and the KSA-XL-BOMD simulations will eventually decorrelate because the system is chaotic – it will just take longer time.

Figure 3.c presents the IR spectra derived from these three Langevin thermostat simulations using the same methodology as applied in the Methanol section. There is a good agreement between direct BOMD and both KSA-XL-BOMD and XL-BOMD, but the latter exhibits more noise at the zero-absorption region. The RMSE between whole normalized BOMD and XL-BOMD spectra is 0.046 vs. 0.026 for KSA-XL-BOMD.

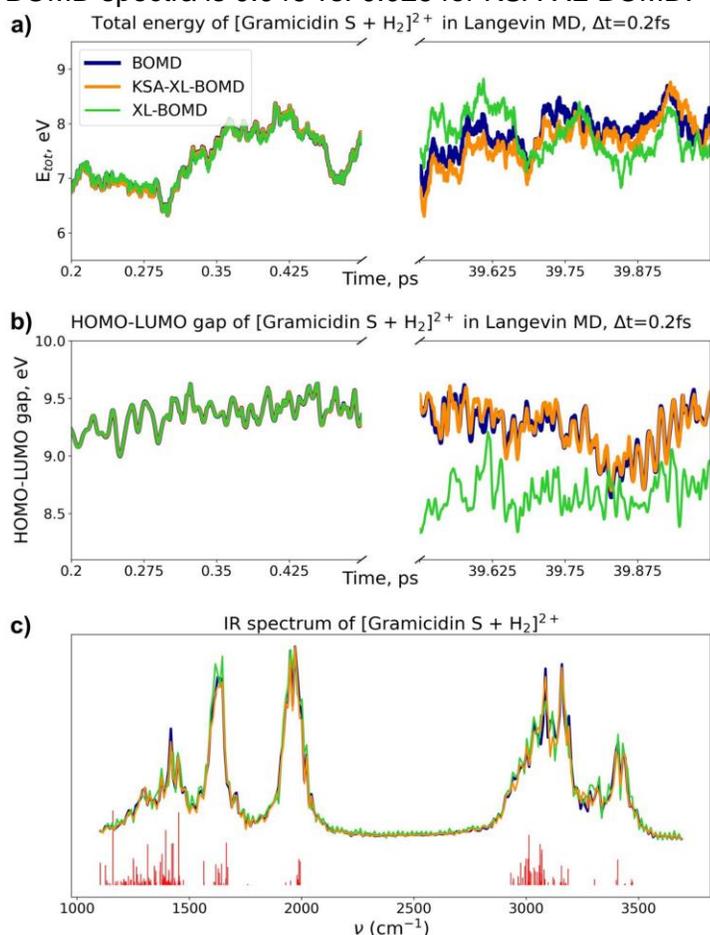

Figure 3. a) Total energy and b) HOMO-LUMO gap in Langevin MD trajectories of $[GS + 2H]^{2+}$. c) IR spectrum of $[GS + 2H]^{2+}$ (gas phase approximation) calculated as a Fourier transform of a dipole moment autocorrelation function (Langevin MD).

The timing per MD step for each method is the following: 0.65 s/step for direct BOMD, 0.45s/step BOMD with DM reused at each MD step, 0.28 s/step for XL-BOMD, and 0.29 s/step for KSA-XL-BOMD. Thus, by a small computational overhead compared to XL-BOMD, KSA-XL-BOMD manages to keep the electronic dynamics close to the direct BOMD baseline in simulations of the 176-atom decapeptide. These results are summarized in Table 1.

**Table 1.** Timing, computational speed up, energy drift, and RMSE of IR spectrum of a doubly protonated Gramicidin S in direct BOMD, BOMD with DM initialization from previous step (reuse DM), KSA-XL-BOMD, and XL-BOMD simulations.

| Method | s/step | Speed up (w.r.t. direct BOMD) | E drift (eV), dt=0.1 fs | E drift (eV), dt=0.7 fs | IR spectrum RMSE (w.r.t. direct BOMD) |
|---|---|---|---|---|---|
| **BOMD (direct)** | 0.65 | -- | -- | -- | -- |
| **BOMD (reuse DM)** | 0.45 | 30.8% | $-3.8 \times 10^{-4}$ | $7.3 \times 10^{-5}$ | -- |
| **Rank-2 KSA-XL-BOMD** | 0.29 | 55.4% | $-2.2 \times 10^{-7}$ | $-7.2 \times 10^{-6}$ | 0.026 |
| **XL-BOMD** | 0.28 | 56.9% | $-6.5 \times 10^{-7}$ | $-2.0 \times 10^{-5}$ | 0.046 |

Computational speed up and RMSE values are given with respect to (w.r.t) direct BOMD simulations

## Fullerenes Test Set

Previously, Zhou *et. al.* have shown that the true strength of PySeQM is its ability to process large molecules on GPUs.[12] Since GPUs are known to perform best on fairly large arrays of data, the timing tests presented below are focused on a set of fullerenes with the relatively large $C_{70}$ molecule being the smallest system in the set. Being homonuclear all-carbon molecules, fullerenes are good systems for testing time scaling of MD methods implemented in PySeQM. Here, we perform MD simulations for a set of neutral fullerenes $C_{70}$, $C_{180}$, $C_{240}$, $C_{320}$, $C_{540}$, $C_{720}$, $C_{840}$ (Figure 4.a) and compare timing of BOMD (re-initializing (labeled *reinit*) and re-using (labeled *reuse*) the DM at each time step), XL-BOMD, and KSA-XL-BOMD (rank 2, 3, 4) on GPU (NVIDIA RTX A6000). The geometries of these molecules are first relaxed using a steep descent optimizer. Then, we perform MD simulations for 1000 steps using Langevin thermostat with a 0.4 fs integration time step and a 300K temperature. The velocities are assigned randomly using the same seed for random number generator in each simulation.

Figure 4.a shows the average time cost per MD step in BOMD, XL-BOMD, and KSA-XL-BOMD. Expectedly, BOMD with DM reinitialization at each step is the most time-consuming approach. Reusing the DM from the previous step helps reduce the time cost. However, it is still ~1000% more expensive than XL-BOMD and ~700% more expensive compared to the rank-2 KSA-XL-BOMD simulation for $C_{840}$ (the largest system under consideration).

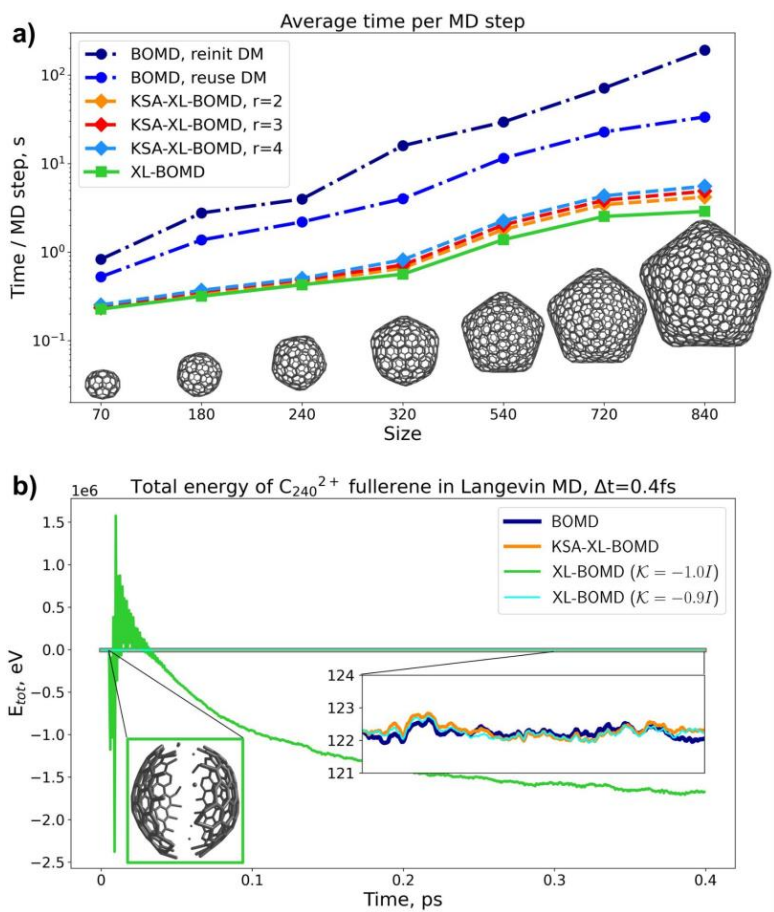

Figure 4. a) Average time spending per MD step for seven neutral fullerenes ($C_{70}$, $C_{180}$, $C_{240}$, $C_{320}$, $C_{540}$, $C_{720}$, $C_{840}$). b) The total energy of $C_{240}^{2+}$ fullerene in Langevin MD. Left inset: structure of $C_{240}^{2+}$ at 20th step in XL-BOMD simulation. Right inset: zoomed total energy in BOMD, rank-2 KSA-XL-BOMD, and XL-BOMD with modified kernel simulations.

KSA-XL-BOMD requires more time per MD step than the original XL-BOMD due to an additional computational overhead associated with the calculation of kernel $\mathcal{K}$, so the rank-2 KSA-XL-BOMD simulation is 45% more expensive than XL-BOMD for the largest system. For the smallest system examined, $C_{70}$, the overhead is only about 3%.

The timings per MD step for the largest $C_{840}$ are the following: 190.7 s/step for direct BOMD with DM reinitialization, 33.5 s/step for BOMD with initializing DM from the previous step, 2.6 s/step for XL-BOMD, and 4.1 s/step for rank-2 KSA-XL-BOMD.

Overall, the rank increase has a moderate effect on the KSA-XL-BOMD timing. Rank-3 and rank-4 are 16% and 33% more time consuming than rank-2 for $C_{840}$. However, for all systems discussed in this work, we did not observe accuracy improvement when increasing the rank from 2 to 3 or 4.

AM1 generally provides larger HOMO-LUMO gaps compared to standard DFT Generalized Gradient Approximation level. For example, the HOMO-LUMO gap of a neutral $C_{240}$ is ~1.1 eV and ~5.2 eV for DFT and AM1 (PySeQM) calculations, respectively. To test the performance of the methods on systems with small HOMO-LUMO gaps, we examine the three MD approaches on a charged fullerene $C_{240}^{2+}$ starting from the geometry of its neutral counterpart.

The HOMO-LUMO gaps of the fullerene cation are 2.4 eV and 1.6 eV at optimized geometries of charged and neutral systems, respectively, at AM1-PySeQM level of theory.

Figure 4.b depicts the total energy of $C_{240}^{2+}$ along the Langevin MD trajectory. As we can see, XL-BOMD exhibits highly unstable dynamics with the immediate system's explosion (see green line and left inset in Figure 4.b), raising the statistical temperature up to $10^6$ K. In the original PySeQM implementation of XL-BOMD, the kernel $\mathcal{K}$ was chosen as a scaled delta function defined as $-c\boldsymbol{I}$ where $\boldsymbol{I}$ is an identity matrix, and the constant $c$ is set to 1. Here, we tuned $c$ manually to improve the stability of the dynamics. Selecting a smaller constant, $c=0.9$, helps to achieve the stability (see cyan line in Figure 4.b).

Meanwhile, rank-2 KSA-XL-BOMD trajectory remains close to the direct BOMD till the end of the simulation and does not require additional optimization of any parameters. This test case illustrates one of the key improvements of KSA-XL-BOMD over the original XL-BOMD method: an improved stability to electronic structures with small HOMO-LUMO gaps that may have unstable charge sloshing.

## SCF Optimization

Finally we illustrate the application of low-rank KSA approach (fixed rank, no preconditioner) to find the electronic ground state density and compare it with established SCF optimization via linear mixing in a semiempirical context. The same initial guess of DM – a diagonal identity matrix – is used in all tests. As a convergence criterion, we use the electronic energy change, $\Delta E$, which should be less than $10^{-7}$ eV. Figure 5.a depicts the number of SCF iterations vs. the electronic energy change of a charged fullerene $C_{380}^{2+}$. In a linear mixing scheme, we use a fixed 10% mixing coefficient (i.e., 10% of the old DM and 90% of the new DM) optimized for the fastest SCF convergence. Adaptive mixing and adaptive mixing with Pulay mixing require more iterations for this particular molecule. Linear mixing requires 139 SCF iterations to achieve $10^{-7}$ eV convergence with the speed of 0.39 s/iteration and 54.21 s of total time on the NVIDIA RTX A6000. Meanwhile, the DM achieves the convergence criterion for 79 iterations within rank-2 KSA scheme with 0.51 s/iteration which is 40.29 s of total time for the whole SCF cycle. Rank-3 KSA gives the converged DM for 36 iterations with a speed of 0.59 s/iteration which takes 21.24 s for the whole SCF cycle. Although there is an additional overhead associated with the low-rank updates of the kernel $\mathcal{K}$ at each SCF iteration, a faster convergence rate makes the KSA approach more efficient for $C_{380}^{2+}$. This timing benefit can, perhaps, be attributed to a nontrivial electronic structure of this unspiralable[45] fullerene, which has an extra charge, high strain at tetrahedral vertices, and a non-relaxed geometry.

For example, the neutral $C_{840}$ fullerene at the optimized geometry (Figure 5.b) took only 51 SCF iterations using a linear mixing optimization (5% mixing coefficient) with a computational speed of 1.80 s/iteration, which results in 91.8 s of total time. Although both KSA rank-2 and rank-3 methods require less iterations to achieve $10^{-7}$ eV convergence, they also require significantly more computational time per iteration: both take 39 iterations with 3.20 s/iteration for rank-2 and 3.89 s/iteration for rank-3. This accounts for a total DM optimization time of 124.8 s and 151.71 s, respectively.

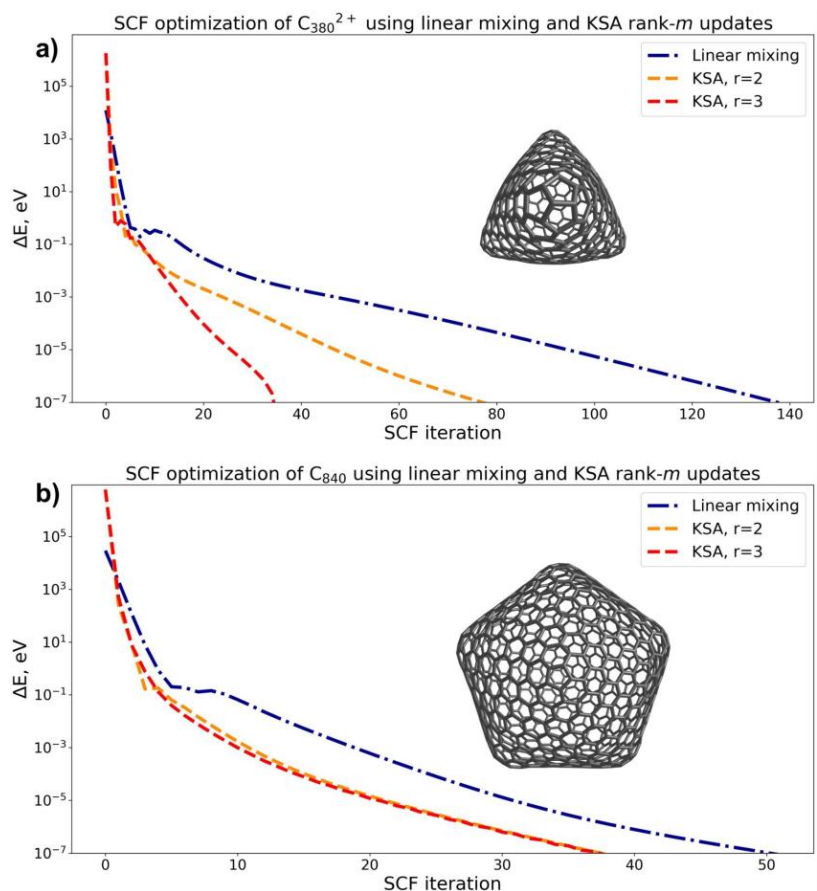

Figure 5. Self-consistent field (SCF) optimization of a) non-relaxed $C_{380}^{2+}$ and b) relaxed $C_{840}$ fullerenes using linear mixing and low-rank Krylov Subspace Approximation (KSA) of the kernel. The rank is denoted as r in the legend. The energy change, $\Delta E$, of $10^{-7}$ eV is used as a convergence criterion.

As we can see, the timing benefit of SCF optimization via low-rank KSA is not systematic. For well-converging electronic structures, linear mixing does SCF optimization faster, although requiring more iterations. Thus, the low-rank KSA approach is not expected to compete with Broyden's class of optimization methods[46–51] for general problems but rather could be used as a complementary tool for systems that exhibit problematic SCF convergence via established schemes. See ref [17] for an extended discussion of a low-rank KSA optimization within SCC-DFTB theory.

# Conclusions

In this work, we present the formulation of Krylov Subspace Approximation for eXtended Lagrangian Born-Oppenheimer Molecular Dynamics (KSA-XL-BOMD) for semiempirical quantum chemistry and interfaced it with semiempirical quantum mechanics Hamiltonians in GPU-accelerated PySeQM package. The new method involves finite thermal Hartree-Fock formalism, canonical density matrix perturbation theory, and an adaptive Krylov subspace approximation for the integration of the electronic equations of motion. The incorporation of a low-rank approximation of a fourth-order kernel in KSA-XL-BOMD results in improved dynamics of the density matrix and enables accelerated MD simulations by avoiding the expensive SCF algorithm.

We exemplify an improved dynamics of density matrices in KSA-XL-BOMD compared to the original XL-BOMD by calculating the IR spectrum of methanol and Gramicidin S decapeptide as a Fourier transform of a dipole moment autocorrelation function. The IR spectrum from KSA-XL-BOMD simulations is almost indistinguishable from the one derived from direct BOMD. Since the dipole moment is directly related to the density matrix, these results indicate that the potential energy surface provided by KSA-XL-BOMD is almost identical to the surface of the time-consuming direct BOMD.

Analysis of the total energy drift in Gramicidin S MD trajectory reveals a higher stability of KSA-XL-BOMD to larger time steps. The test case of a charged fullerene $C_{240}^{2+}$ highlights the KSA-XL-BOMD stability when applied to systems with small HOMO-LUMO gaps. Furthermore, the analysis of total energies and HOMO-LUMO gaps in MD trajectories of Gramicidin S and $C_{240}^{2+}$ fullerene using Langevin thermostat provides some evidence that KSA-XL-BOMD keeps the electronic dynamics in the closest proximity to direct BOMD.

Timing tests on a set of large fullerenes illustrate the ability of KSA-XL-BOMD in the current PySeQM implementation to perform fast and accurate simulations of large molecules. The timings per MD step for the largest $C_{840}$ are the following: 190.7 s/step for direct BOMD with DM reinitialization, 33.5 s/step for BOMD with initializing DM from the previous step, 2.6 s/step for XL-BOMD, and 4.1 s/step for rank-2 KSA-XL-BOMD. Thus, there is a moderate computational overhead in comparison with the original XL-BOMD formulation. This increase is associated with the calculation of a low-rank approximation of a fourth-order kernel $\mathcal{K}$. However, the KSA-XL-BOMD is still faster than direct BOMD by one to two orders of magnitude.

Apart from MD simulations, we demonstrate the ability of low-rank KSA to perform the SCF optimization of a density matrix. Although the presented scheme is not expected to compete with conventional methods, such as Pulay or Anderson mixing, for general problems, it could be used as a complementary approach for systems that are particularly hard to converge with established methods.

Overall, the KSA-XL-BOMD in the current PySeQM implementation can treat systems with several thousands of atoms and with elements from H to Cl, which cover all organic molecules. PySeQM also supports metalloids and nonmetals up to I. While the current release of PySeQM can only treat closed-shell systems supports *s* and *p* orbitals, future developments will augment this package with other semiempirical models such as PM6 and OMx Hamiltonians, adding the support of *d* orbitals and explicit treatment of orbital overlaps S. Other potential improvements include open shell systems, excited states calculator, and excited state dynamics beyond Born-Oppenheimer approximation, which will enable a large variety of simulations including photophysics and catalysis modeling. As was shown previously, semiempirical methods in PySeQM can be significantly boosted in accuracy to the level of *ab-initio* DFT accuracy via a machine-learned (ML) fit of Hamiltonian parameters.[12] Combining KSA-XL-BOMD with ML is another topic of further research.

PySeQM source code with example scripts and cartesian coordinates of structures discussed in this work are freely available at https://github.com/kulichenko-LANL/PYSEQM_dev/tree/main.

# Acknowledgements

M.K. and N.F. acknowledge support from the Los Alamos National Laboratory (LANL) Directed Research and Development (LDRD) funds and financial support from the Director's Postdoctoral Fellowship at LANL. K.B., S.T., and B.N. acknowledge support from the US DOE, Office of Science, Basic Energy Sciences, Chemical Sciences, Geosciences, and Biosciences Division under Triad National Security, LLC ("Triad") contract Grant 89233218CNA000001 (FWP: LANLE3F2).

A.N. acknowledges support by the U.S. Department of Energy Office of Basic Energy Sciences, Chemical Sciences, Geosciences, and Biosciences Division (FWP: LANLE8AN "Next generation quantum-based molecular dynamics").

This work was performed in part at the Center for Nonlinear Studies (CNLS) and the Center for Integrated Nanotechnology (CINT), a US Department of Energy (DOE) and Office of Basic Energy Sciences user facility. This research used resources provided by the LANL Institutional Computing Program, which is supported by the US DOE National Nuclear Security Administration under Contract 89233218CNA000001. We also acknowledge the CCS-7 Darwin cluster at LANL for additional computing resources.